\documentclass[useAMS,usenatbib]{mn2e}
\usepackage{url}
\usepackage{graphicx}
\usepackage{verbatim}
\usepackage[usenames]{color}
\usepackage[flushleft]{threeparttable}
\DeclareGraphicsExtensions{.pdf,.png,.jpg,.mps,.eps,.ps}
\usepackage{amsmath}
\usepackage{natbib}
\usepackage{color}
\usepackage{graphicx}
\usepackage{lscape}
\usepackage{gensymb}

\newcommand{\quotes}[1]{``#1''}
\newcommand\nustar{{\it NuSTAR}}

\newcommand\asca{{\it ASCA}}
\newcommand\ginga{{\it Ginga}}
\newcommand\sax{{\it BeppoSAX}}
\newcommand\chandra{{\it Chandra}}

\newcommand\suzaku{{\it Suzaku}}
\newcommand\rxte{{\it RXTE}}

\newcommand\exosat{{\it EXOSAT}}

\newcommand\kev{{\rm~keV}}

\newcommand\kms{\ifmmode {\rm~km\ s}^{-1} \else ~km s$^{-1}$\fi}
\newcommand\Hunit{\ifmmode {\rm~km\ s}^{-1}\ {\rm Mpc}^{-1}
        \else ~km s$^{-1}$ Mpc$^{-1}$\fi}
\newcommand\ctssec{\ifmmode {\rm~count\ s}^{-1} \else ~count s$^{-1}$\fi}
\newcommand\ergsec{\ifmmode {\rm~erg\ s}^{-1} \else
        ~erg s$^{-1}$\fi}
\newcommand\funit{\ifmmode {\rm~erg\ s}^{-1}\;{\rm cm}^{-2} \else
        ~ergs s$^{-1}$ cm$^{-2}$\fi}
\newcommand\phflux{\ifmmode {\rm~photon\ s}^{-1}\;{\rm cm}^{-2}
        \else   ~photon s$^{-1}$ cm$^{-2}$\fi}
\newcommand\efluxA{\ifmmode {\rm~erg\ s}^{-1}\;{\rm cm}^{-2}\;{\rm
        \AA}^{-1} \else ~erg s$^{-1}$ cm$^{-2}$ \AA$^{-1}$\fi}
\newcommand\efluxHz{\ifmmode {\rm~erg\ s}^{-1}\;{\rm cm}^{-2}\;{\rm
        Hz}^{-1} \else ~erg s$^{-1}$ cm$^{-2}$ Hz$^{-1}$\fi}
\newcommand\cc{\ifmmode {\rm~cm}^{-3} \else cm$^{-3}$\fi}
\newcommand\FWHM{\ifmmode {\rm~FWHM} \else ${\rm~FWHM}$\fi}
\newcommand\Msun{\ifmmode M_{\odot} \else $M_{\odot}$\fi}
\newcommand\Lsun{\ifmmode L_{\odot} \else $L_{\odot}$\fi}

\newcommand\hbeta{\ifmmode {\rm H}\beta \else H$\beta$\fi}
\newcommand\Kalpha{\ifmmode {\rm K}\alpha \else K$\alpha$\fi}
\newcommand\nh{\ifmmode N_{\rm H} \else N$_{\rm H}$\fi}
\usepackage{psfrag}
\usepackage{graphicx}
\usepackage{url}
\usepackage{verbatim}
\usepackage{color}

\title[\nustar{} spectroscopy of Cyg~X--2]{\nustar{} view of the Z-type neutron star low-mass X-ray binary Cygnus~X--2}
\author[Mondal et al.]{\parbox[]{6.5in}{{Aditya S. Mondal$^{1}\thanks{E-mail: adityas.mondal@visva-bharati.ac.in}$, G. C. Dewangan$^{2}$
, M. Pahari$^{2}$, B. Raychaudhuri$^{1}$ }  \\
\small
$^{1}$Department of physics, Visva-Bharati, Santiniketan, West Bengal-731235, India \\
$^{2}$Inter-University Centre for  Astronomy \& Astrophysics (IUCAA), Pune, 411007 India \\
}}

\date{\today}
\begin{document}
\maketitle
\begin{abstract}
We report on the \nustar{} observation of the Z-type neutron star low-mass X-ray binary Cygnus X--2 performed on 7 January 2015. During this observation, the source exhibited a sudden decrease in count rate (dips) and stronger variability in $3-79\kev$ X-ray lightcurve. The hardness-intensity diagram shows that the source remained in the so-called \quotes{normal branch} of the Z-track, although an extended \quotes{flaring branch} is observed during the dips. The source was in a soft spectral state with the $3-45\kev$ luminosity of $L\simeq(0.5-1.1)\times 10^{38}$ erg s$^{-1}$, assuming a distance of 8 kpc.  Both the non-dip and dip X-ray spectra are well represented by models in which the soft band is dominated by the emission from the disc, while the hard X-ray band is dominated by the Comptonized emission from the boundary layer/corona and its reflected emission from the disc. The X-ray spectrum also revealed a broad Fe K$\alpha$ emission line which is nearly symmetric at the higher flux and asymmetric when the flux is reduced by a factor of $\sim 2$. The relativistic reflection model predicts the inner radius of the accretion disc as $R_{in}\simeq2.5-6.0\:R_{ISCO}\:(\simeq30-73$ Km) for the non-dip state and $R_{in}\simeq2.0-2.6\:R_{ISCO}\:(\simeq24-32$ Km) for the dip state. If the inner disc is truncated due to the pressure arising from a magnetic field, this implies an upper limit of the magnetic field strength of $\leq7.6\times10^{9}$ G at the magnetic poles which is consistent with other estimates.       
\end{abstract}
\section{introduction}
An X-ray binary consists of a compact star (white dwarf, neutron star or black hole) and a companion star/donor star (may be a main sequence star or an evolved star). If the compact object in an X-ray binary system is a neutron star (NS) and the companion star is a low-mass star ($\le1M_{\odot}$), it is called a neutron star low mass X-ray binary (NS LMXB). In NS LMXB systems, the low-mass star loses mass which is accreted onto the weakly magnetized NS. NS LMXBs are usually classified into two sub-classes  --  the "Z" and "atoll" sources, on the basis of their correlated X-ray spectral and fast timing behavior \citep{1989A&A...225...79H}. Z source traces out a Z-shape pattern in the X-ray colour-colour diagram (CD) on time scales of hours to days. The branches of the Z pattern are called, from top to bottom, the horizontal branch (HB), the normal branch (NB) and the flaring branch (FB). Through the branches, the overall intensity varies smoothly on time scales of weeks to months. The transition between the HB and the NB is called the hard vertex, and between the NB and the FB the soft vertex. The variations of the mass transfer rate to the compact object are thought to produce the movement along the Z track (e.g. \citealt{1989A&A...225...79H, 1996A&A...311..197K,  1997A&A...323..399W, 1991ASPC...20..299L}). Z sources can further be divided into two groups, the Cyg-like sources (e.g. Cyg~X--2, GX 5-1, GX 340+0) and the Sco-like sources (Sco X-1, GX 349+2, GX 17+2). Cyg-like sources display significant motion of their Z pattern in the CD, whereas the Sco-like sources do not. Usually, Cyg-like sources show all three branches of the Z track while for Sco-like sources the HB is weak but FB is much stronger \citep{2012MmSAI..83..170C}. Theoretical studies \citep{1995ApJ...454L.137P} indicate that the Cyg-like sources are associated with stronger magnetic fields ($B\sim5\times10^{9}$ G) than the Sco-like sources ($B\sim10^{9}$ G). Similarly, in Atoll sources the accretion rate increases from the so-called island state to the upper banana branch.  \\

Cygnus X--2 (Cyg~X--2) is a well studied bright low-mass X-ray binary which is classified as a Z source as it traces out a Z-shape pattern in the X-ray CD (e.g. \citealt{1989A&A...225...79H, 1996A&A...311..197K}) . It was discovered by \citet{1966AJ.....71..379B} with a sounding rocket experiment. Cyg~X--2 was first observed by \exosat{} for a continuous time stretch of $14$ h starting on July 23, 1984, 02:07 UT with the help of Gas scintillation proportional counter \citep{1981SSRv...30..525P} and one half of the medium energy detectors \citep{1981SSRv...30..513T}. The companion of Cyg~X--2 is an evolved, late-type star of a mass ranging between $0.4 - 0.7 M_{\odot}$, the spectral type of which seems to vary from A5 to F2 \citep{1979ApJ...231..539C}. The compact object was identified as an NS after the observation of thermonuclear X-ray bursts \citep{1984ApJ...281..826K, 1998ApJ...498L.141S} in the X-ray lightcurve of Cyg~X--2. \citet{2002ApJ...570L..25T} used the \rxte{} burst data to estimate the NS mass to be about $1.4M_{\odot} $ and the radius to be about 9~km. Moreover, \citet{1998ApJ...493L..39C} calculated a neutron star mass of $1.71\pm0.21M_{\odot} $, while \citet{2009MNRAS.395.2029E} derived a mass of $1.5\pm0.3M_{\odot}$. The source itself has an estimated distance of $8 - 11$ kpc \citep{1979ApJ...231..539C, 1998ApJ...498L.141S}. Optical observation estimated the source distance to be $7.2\pm1.1$ kpc \citep{1999MNRAS.305..132O}. \citet{1979ApJ...231..539C} measured the mass limits of the two components of this binary system and also an orbital period of $\sim9.8$ days along with other orbital parameters. Recent high-resolution optical spectroscopy by \citet{2010MNRAS.401.2517C} also provided a refined orbital solution with a period of $9.84450 \pm 0.00019$ days. The first $160$ days of the \rxte{} \citep{1993A&AS...97..355B} All Sky Monitor (ASM) data of Cyg~X--2 suggested an $78$ day period in the long-term variations \citep{1996ApJ...473L..45W}. This was also supported by the archival data from {\it Ariel 5} and {\it Vela 5B}. \\

The X-ray spectrum of Cyg~X--2 has been studied extensively over the years. The broad band X-ray spectrum of this source has been studied with \sax{} \citep{1999A&AS..138..399F,2002A&A...386..535D, 2002ApJ...567.1091P}. The spectral analysis using the data from \asca{} and \sax{} confirmed the presence of lines from various K-shell ions \citep{1997A&A...323L..29K, 2002A&A...386..535D}. Like other bright LMXBs, Cyg~X--2 also showed the presence of Fe K$\alpha$ emission in its X-ray spectra. H-like Fe xxvi line emission was found by \citet{1993ApJ...410..796S, 1994AAS...18510204S} using the higher resolution spectrometer of BBXRT. \citet{2002A&A...386..535D} performed the detailed spectral analysis of the source using the \sax{} data along the Z-spectral pattern. They fitted the broad band continuum using a two component model, consisting of a disk blackbody and a Comptonized component and detected the broad emission line features at $1\kev$ and at $6.5-6.7\kev$. \citet{2009ApJ...699.1223S} reported the red-skewed iron line profile from the \suzaku{} spectrum of this source. They attributed this as a result of the line formation in the extended wind/outflow configuration instead of the reflection of the X-ray radiation from a cold accretion disk. \citet{2002MNRAS.331..453D} successfully fitted the \ginga{} data with a reflection model where a Comptonized component irradiates the accretion disk. \citet{2010ApJ...720..205C} also inferred a broad Fe emission line from the \suzaku{} spectra of Cyg~X--2. Many observations of Cyg~X--2 indicated an extended accretion disk corona (ADC). \citet{1988ApJ...329..276V} suggested that high and low flux states reflect changes in the geometrical and optical thickness of the accretion disk and ADC. An evidence of an extended ADC was further supported by the \chandra{} high-resolution grating results \citep{2009ApJ...692L..80S} on highly excited H-like lines of Ne, Mg, Si, S and Fe in Cyg~X--2. The source has been known to exhibit extensive dipping activity in X-ray intensity on the FB. Sometimes hard X-ray tail has been observed in the X-ray spectra of Cyg~X--2 \citep{2002ApJ...567.1091P} which could be due to the presence of a high-energy non-thermal emission. \\

Accurate timing studies, mainly performed with proportional counter array (PCA) on board the \rxte{} satellite, have shown the presence of rapid variability phenomena such as -- quasi-periodic oscillations (QPOs), band limited noise etc. in the frequency range extending from Hz to kHz \citep{1998ApJ...493L..87W}. Thermonuclear X-ray bursts \citep{1984ApJ...281..826K, 1998ApJ...498L.141S} have been observed in the X-ray lightcurve of Cyg~X--2. \citet{1998ApJ...493L..87W} reported the simultaneous detection of twin kHz peaks at $500$ and $860$ Hz and highest single kHz QPO at $1007$ Hz. The source exhibited $18-50$ Hz QPOs in the HB and $5.6$ Hz QPOs in the normal branch \citep{1986Natur.319..469H, 1986ApJ...308..655E,1997A&A...323..399W, 1999MNRAS.308..485K}.  \\

In this paper, we report on the X-ray spectral and timing analysis of high sensitivity, broad bandpass \nustar{} observation \citep{2013ApJ...770..103H} of Cyg~X--2. We particularly focus on the broad relativistic iron line in the $\sim5-8\kev$ band and constrain the stellar radius and/or inner accretion disk radius. In addition, our result confirms the position of the source in the Z-track during this observation. We establish some new results along with reconfirmation of some earlier findings. We organized the paper in the following way. First, we describe the observations and the details of data reduction in sec .2. In sec. 3 and sec. 4, we describe the temporal and spectral analysis, respectively. In sec.5, we discuss our findings and we summarize our results in sec.6.

\section{observation and data reduction}
The source Cyg~X--2 was observed with \nustar{} on 2015 January 7 (MJD $5702.9146$) for a total exposure time of $\sim 23.7$ ks (Obs. ID: $30001141002$). \nustar{} data were collected with the two  co-aligned grazing incidence hard X-ray imaging telescopes (FPMA and FPMB) in the $3-79 \kev$ band.

We reprocessed the data using the standard \nustar{} data analysis software ({\tt NuSTARDAS v1.5.1}) and {\tt CALDB} ($v20150904$). We used the {\tt nupipeline}  (version v 0.4.3) to filter the event lists. We used a circular region with a radius of 100 arcsec to extract the source events. We also extracted background events from  the corner of the detectors deviod of any sources using circlular region of similar size as the source region. We employed the task {\tt nuproduct} to create lightcurve, spectra and response files for the FPMA and FPMB. We grouped the FPMA and FPMB spectral data with a minimum of 100 counts per channel, and  fitted the two spectra simultaneously. 

\begin{figure}
   \centering
   \includegraphics[width=8.0cm, angle=0]{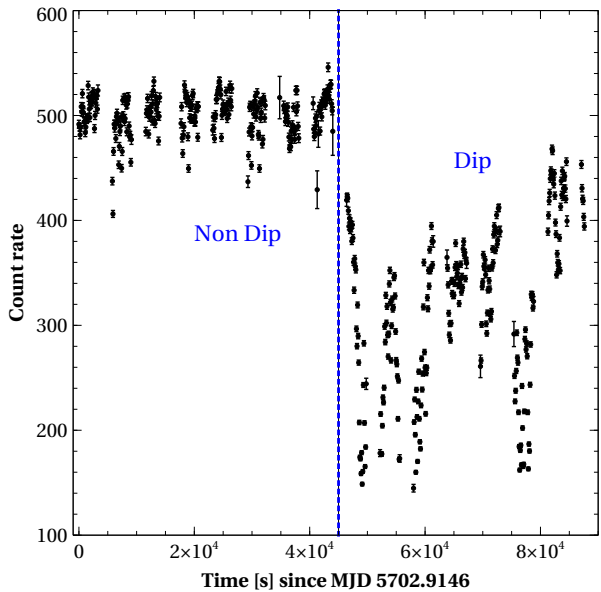}
   \caption{The background corrected, full band ($3-79\kev$) \nustar{}/FPMA lightcurve of the source Cyg~X--2 with $100$ s bins.} 
   \label{Fig1}
   \end{figure}
 
\section{Temporal Analysis}
\subsection{Light curve}
In Fig.~\ref{Fig1}, we show the $3-79\kev$ \nustar{} FPMA light curve of Cyg~X--2 spanning $\sim 1$ day and binned with 100~s. 
The X-ray lightcurve clearly shows a reduction in intensity from a count rate of $\sim 500$ counts s$^{-1}$ in the first half to $\sim 280$ count s$^{-1}$ in the second half of the observation. The drop in count rate occured in $\sim 5\times 10^{4}$ s. In addition, a number of dips in the second half of the observation with decrease in count rates by about a factor of $\sim 2$ are clearly seen. These variabililty events appear as absorption dips as observed for many dipping class of LMXBs. 

\subsection{Hardness ratio and Hardness-intensity diagram}
 
\begin{figure}
   \centering
   \includegraphics[scale=0.35, angle=-90]{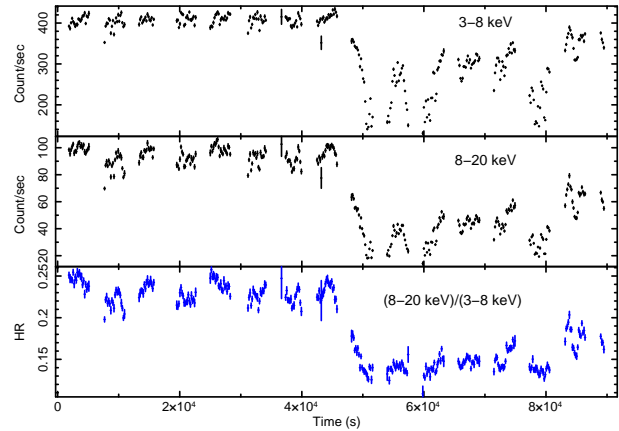}
   \caption{The upper and the middle panels shows the source count rate in the energy band $3-8$ keV and $8-20$ keV, respectively. Bottom panel shows the ratio of count rate in the energy band $8-20$ keV and $3-8$ keV. } 
  \label{Fig2}
   \end{figure}

To further investigate the variability of Cyg~X--2, we extracted the lightcurves in the $3-8\kev$ and the $8-20\kev$  bands with 100~s bins. These two lightcurves are shown in the upper and middle panel of Fig.~\ref{Fig2}, respectively. The lower panel of Fig.~\ref{Fig2} shows the hardness ratio which is defined as the ratio of counts per second in the $8-20\kev$ to that in the $3-8\kev$ band. The hardness ratio shows a factor of $\sim 1.6$ softening of the spectrum during the dips. This behavior is typical and observed for many dipping LMXB sources. The significant hardening in the observed X-ray spectrum indicates a change in the underlying physical conditions. 

\begin{figure}
   \centering
   \includegraphics[width=8.0cm, angle=0]{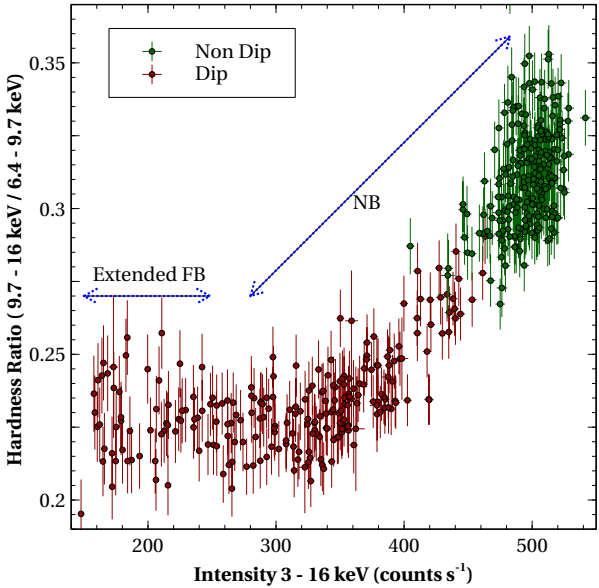}
   \caption{The hardness-intensity diagram (HID) of Cyg~X--2. Intensity is taken as $3-16$ keV source count rate and hardness ratio has been taken as the ratio of count rate in the energy band $9.7-16$ keV and $6.4-9.7$ keV. The variation of hardness ratio with intensity for the non-dip and dip segments of the lightcurve are shown in green and red colours, respectively.   } 
   \label{Fig3}
   \end{figure}

We further examined the spectral variation of the source by plotting the \nustar{}  hardness-intensity diagram (HID) which is shown in Fig.~\ref{Fig3}. We calculated the hardness ratio as the count rate ratio between $9.7-16\kev$ and $6.4-9.7\kev$ while the intensity is defined as the count rate in the $3-16 \kev$ band. We used the same definition of hardness ratio and intensity as that of \citet{2001MNRAS.321..537W} and \citet{2002ApJ...567.1091P} to compare with the HID of \rxte{}/\sax{} data. It is seen that the HB is not visible in the HID and the source moved along the Normal Branch (NB) of its Z-track. Thus, comparing the HID with earlier work by \citet{2002ApJ...567.1091P} and \citet{2001MNRAS.321..537W}, we confirmed that the source remained in a particular branch, namely the NB, during the \nustar{} observation. Although a little deviation from the normal branch has been observed at the lower intensity ($\sim200$ counts s$^{-1}$). This might be the transitions between the NB and the FB or an extended FB (according to \citealt{2001MNRAS.321..537W}).  \\ 

\subsection{Power density spectra}
\begin{figure*}
   \includegraphics[scale=0.35, angle=-90]{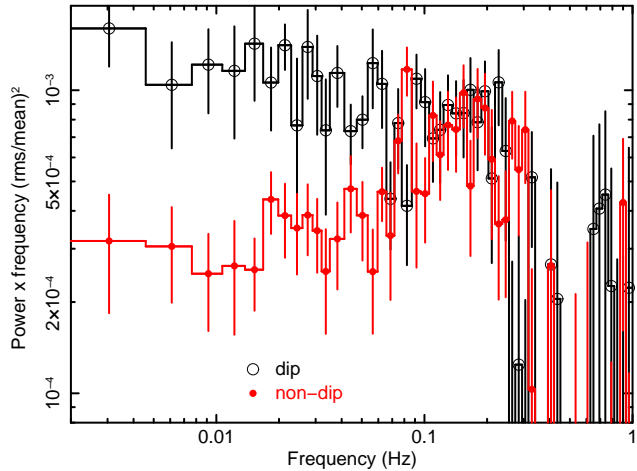}
   \caption{PDS obtained after taking 3 ks continuous segments of the non-dip and dip lightcurve using \nustar{}/FPMA data in the energy band $3.0 - 78.0 \kev$ with a binning of $100$ ms.} 
  \label{Fig4}
   \end{figure*}

We extracted power density spectra (PDS) from both the non-dip and the dip part of the \nustar{}/FPMA light curve in the energy band $3-78 \kev$. Light curves from both the non-dip and dip part consist of one $\sim 3$ ks continuous stretch with a bin time of $100$ ms. The extracted PDS are rms-normalized and Poisson noise subtracted. PDS from the continuous part of the lightcurve from both the non-dip and dip part are shown in Fig.~\ref{Fig4}. Here, we observe that at the low frequency (below $0.1$ Hz) fractional rms variability during the dip part is higher by few factors than the non-dip part fractional variability. Therefore, long-time scale ($10$ s or more) variability increases as we move from NB to the extended FB. The increase in fractional rms as we move from the non-dip (NB) to the dip (extended FB) state is consistent with the fact that the higher rms variability at long time scale is expected from the inner disk as it moves closer to the neutron star surface.        

\begin{figure}
   \centering
   \includegraphics[width=7.0cm, angle=0]{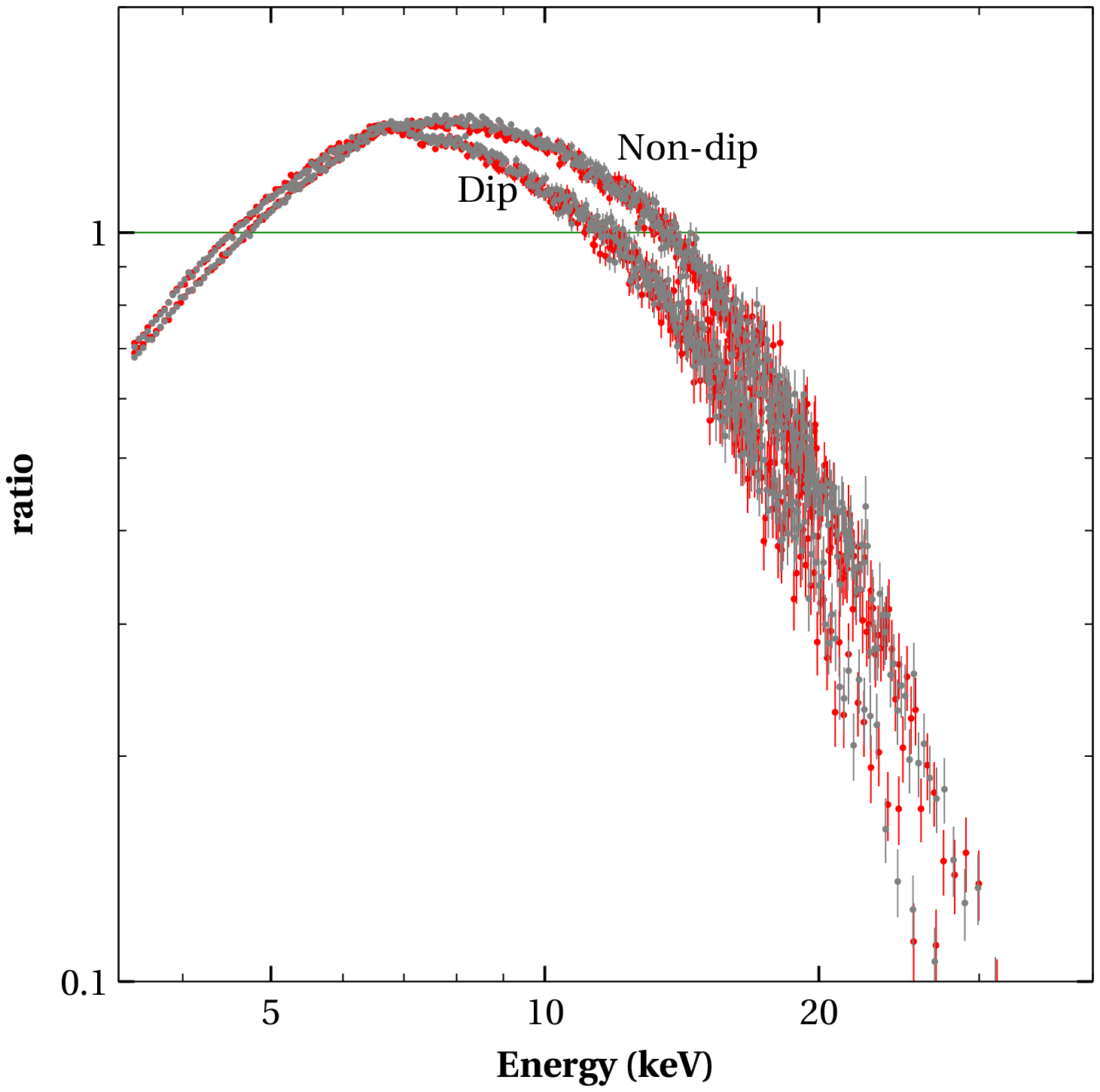}
   \caption{ Data to model ratio as a function of energy as observed from \nustar{} FPMA/FPMB spectra of the source Cyg~X--2 in the time interval of non-dip and dip when both the spectra are fitted by an absorbed power-law model.} 
   \label{Fig5}
   \end{figure}

\section{spectral analysis}
We extracted the FPMA/FPMB spectra after selecting the time intervals and intensities from the \nustar{} lightcurve corresponding to non-dip and dip durations (labelled as non-dip and dip in Fig.~\ref{Fig1}). A non-dip spectrum was selected from the part of the \nustar{} lightcurve spanning  $0-45$ ks with  intensity $> 400$ counts s$^{-1}$. The dip spectrum was extracted from the interval $46-86$ ks with intensity $< 380$ counts s$^{-1}$. For the non-dip as well as the dip states, we simultaneously fitted \nustar{}/FPMA and FPMB spectral datasets using {\tt XSPEC v 12.8.2} \citep{1996ASPC..101...17A}. We tied the parameters of the two datasets but introduced a constant factor to account for cross normalization between the two detectors. The constant factor was fixed to 1 for \nustar{}/FPMA and kept free for FPMB. We noticed that the source was detected significantly above the background up to $\simeq45$ \kev. We quote the uncertainties on model parameters at the $90\%$ confidence level.  

To begin with, we fitted both the non-dip and dip spectrum separately with a absorbed power-law model. For both the spectra, power-law photon index and the normalization take the values of $\Gamma\sim 3$ and $\sim70\;\text{ph}\;\text{cm}^{-2}\;\text{s}^{-1}\;\text{keV}^{-1}$, respectively. We show data to model ratio as a function of energy in Fig.~\ref{Fig5}. Both the spectra show the presence of some excess in the data to model ratio plot $\sim5-8\kev$  and there may exist a possible cut-off around $10-20 \kev$ which are the typical characteristics of the soft state, although no conclusion can be drawn from this plot alone.

\begin{figure*}
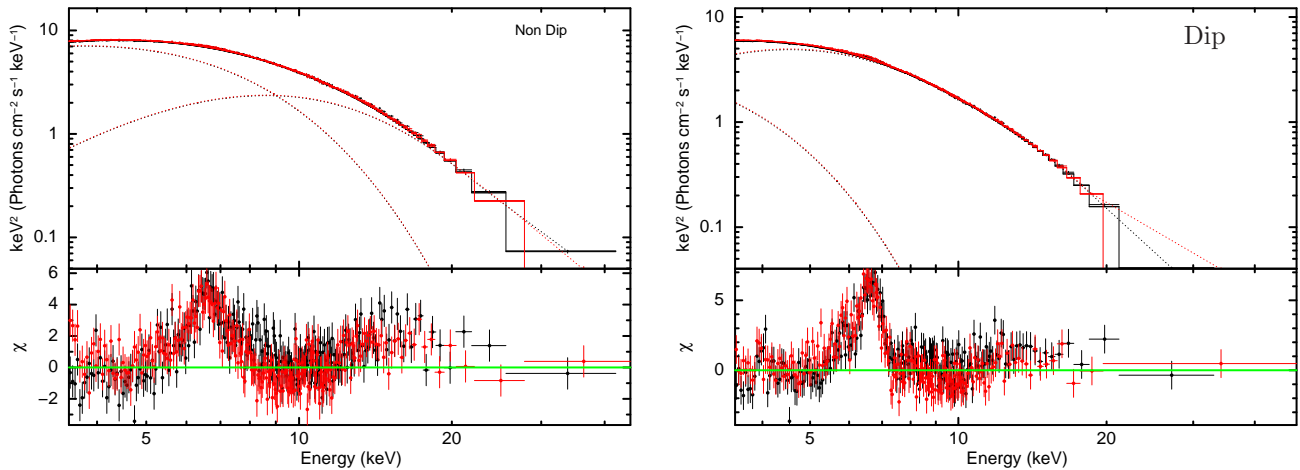

\psfrag{ND}{\large Non-Dip}
\psfrag{Dip}{\large Dip}
   \includegraphics[scale=0.35, angle=-90]{fig6a.ps}
   \includegraphics[scale=0.35, angle=-90]{fig6b.ps}
  \caption{\nustar{} (FPMA in black, FPMB in red) unfolded spectra. The spectral data were rebinned for visual clarity. Continuum is fitted with the model consisting of a disk blackbody and Comptonized emission. Model used: {\tt TBabs$\times$(diskbb+compTT)}. It revaled un-modelled broad emission line $\sim 6-7$ keV as well as the higher energies. } 
  \label{Fig6}
   \end{figure*} 

\subsection{Continuum modeling}
We used a continuum spectral model consisting of a soft emission component from the disk ({\tt diskbb} in {\tt XSPEC}) and a thermal Comptonization component {\tt compTT} \citep{1994ApJ...434..570T}, modified by the interstellar absorption modelled by {\tt tbabs} with {\tt vern} cross section \citep{1996ApJ...465..487V} and {\tt wilm} abundances \citep{2000ApJ...542..914W}. This two-component model often used by many authors to fit the broadband continuum of many NS LMXBs. The {\tt compTT} model describes the Comptonization of soft photons in a hot plasma (characterised by temperature, $kT_{e}$ and optical depth, $\tau$), and results in spectra with significant curvature close to the plasma temperature. Our continuum model, {\tt tbabs$\times$(diskbb+CompTT)}, resulted in $\chi^{2}/dof$=$1523/920$ and $1601/770$ for the non-dip and dip spectrum, respectively (where $dof$ is the degrees of freedom). The parameters from the continuum spectral fits for the non-dip and the dip spectrum are given in Table~\ref{continuum}. \\

It should be noted that for the Z-track sources, the high energy cut-off is relatively low ($\sim5\kev$) which does not allow to determine the powerlaw photon index ($\Gamma$) properly \citep{2010A&A...512A...9B}. In LMXBs the X-ray spectrum above $6\kev$ or so is typically modelled as either a thermal Comptonization or a hot black body. When we tried to fit the spectra with the combination of a multi-temperature disk black body ({\tt diskbb}) and a single temperature black body ({\tt bbody}) component, an excess of counts above $\sim 30 \kev$ was apparent in both the spectra. As the spectra of Z sources sometimes shows the presence of hard tail, we added a power-law component to fit this. This combination of spectral models, {\tt tbabs$\times$(diskbb+bbody+powerlaw)}, is also frequently used for the soft state spectra of many NS LMXBs \citep{2007ApJ...667.1073L, 2010ApJ...720..205C, 2013ApJ...779L...2M}. However, such a model does not provide satisfactory fit with $\chi^{2}/dof$=$1545/920$ and $1650/770$ for the non-dip and dip spectra, respectively. The best-fitting continuum parameters for both the spectra are shown in Table~\ref{continuum}. \\

 In Fig~\ref{Fig6}, we show the fitted continuum model {\tt tbabs$\times$(diskbb+CompTT)} and the $\chi$ residuals. Clearly, there are significant residuals near $6-7\kev$ in both spectra and an excess flux in the $12-20\kev$ band which is prominent for the non-dip spectrum compared to the dip spectrum. The broad residuals $\sim6-7\kev$ indicate the presence of a broad iron $K\alpha$ line. Therefore, we added a {\tt Gaussian} line to model the excess seen $\sim6-7\kev$. The improvement of the fit was statistically significant with $\Delta\chi^{2}=-452$ and $-710$ for 3 additional parameters for the non-dip and dip spectrum, respectively. The line is centered at $\sim6.44\kev$ with a width $\sigma\sim0.96\kev$ for the non-dip spectrum. For the dip spectrum the width of the line was found to be $\sigma\sim0.37\kev$ and the center at $\sim6.55\kev$. However, {\tt Gaussian} model fails to provide satisfactory fit.  \\

\begin{figure*}
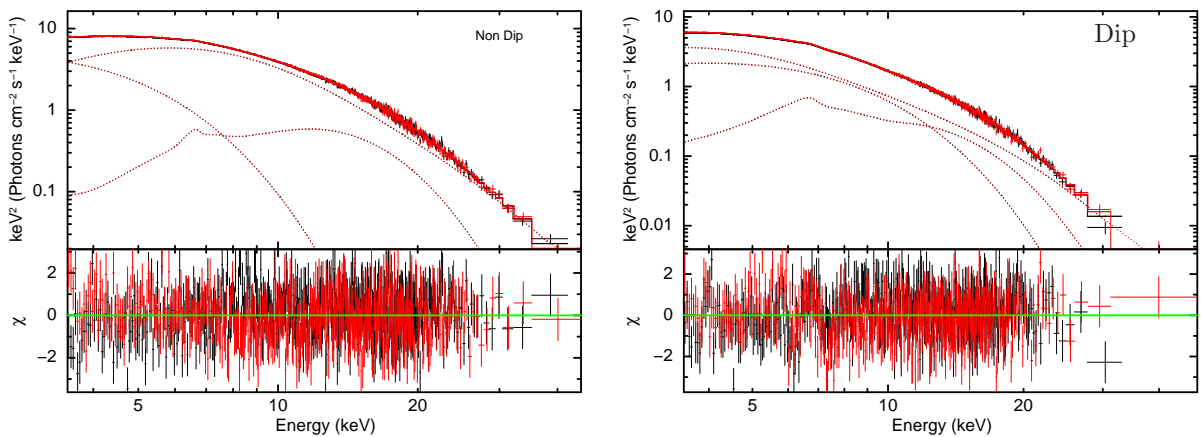

\psfrag{ND}{\large Non-Dip}
\psfrag{Dip}{\large Dip}
   \includegraphics[scale=0.32, angle=-90]{fig7a.ps}
   \includegraphics[scale=0.32, angle=-90]{fig7b.ps}
 \caption{The \nustar{} (FPMA in black, FPMB in red) unfolded spectra of Cyg~X--2 in the non-dip (left) and dip states (right) and the best-fitting fitted model consisting of a disk blackbody, Comptonized emission and relativistically blurred reflection i.e.,{\tt TBabs$\times$(diskbb+compTT+relconv*reflionx)}. Lower panel shows residuals in units of $\sigma$.} 
   \label{Fig7}
   \end{figure*}

\begin{table}
\centering
\caption{Parameters of the continuum models for the non-dip and the dip spectrum. 1$\sigma$ error on all relevant parameters are quoted.} 
\begin{tabular}{|p{2.2cm}|p{1.5cm}|p{1.5cm}|}
\hline
\multicolumn{2}{c}{Model:{\tt TBabs$\times$(diskbb+compTT)}} \\
\cline{1-3} 
Parameter & Non Dip & Dip \\ 
\hline
$kT_{in}$($\kev$) & $1.57_{-0.19}^{+0.08}$ & $1.28_{-0.12}^{+0.06}$  \\
$kT_{e}$($\kev$) & $4.45_{-0.70}^{+7.84}$ &  $2.42_{-0.04}^{+0.11}$ \\
$kT_{seed}$($\kev$) & $2.03_{-0.32}^{+0.23}$ & $1.11_{-0.09}^{+0.05}$  \\
$\tau$   & $1.90_{-1.45}^{+0.74}$  &  $5.47_{-1.20}^{+1.00}$    \\
\hline
\multicolumn{2}{c}{Model:{\tt TBabs$\times$(diskbb+bbody+powerlaw)}} \\
\hline
$kT_{in}$($\kev$) & $1.76\pm0.01$ & $1.42\pm0.01$  \\
$kT_{bb}$($\kev$) & $2.63\pm0.05$ & $2.26\pm0.04$  \\
$\Gamma$ &$3.41\pm0.12$ & $2.05_{-1.70}^{+1.39}$    \\
\hline
\end{tabular}\label{continuum} 
Note: $kT_{in}$= inner disk temperature, $kT_{e}$= electron temperature of the Comptonizing plasma, $kT_{seed}$= seed photon temperature
and $kT_{bb}$= blackbody temperature.
\end{table}

\subsection{Reflection Model}
Fig.~\ref{Fig6} shows clear signature of disk reflection, a broad feature between $5-8\kev$ consistent with Fe K$\alpha$ emission and a flux excess in the $10-20\kev$ consistent with a Compton back-scattering hump. So, we replaced the {\tt Gaussian} component with a more physical model {\tt reflionx} \citep{2005MNRAS.358..211R}. Here we used a modified version of the {\tt reflionx} \footnote{https://www-xray.ast.cam.ac.uk/\~{}mlparker/reflionx\_models \\/reflionx\_alking.mod} that assumes a black body input spectrum illuminates an accretion disk and produces the reflection spectrum (see e.g. \citealt{2010ApJ...720..205C, 2016ApJ...819L..29K, 2016MNRAS.456.4256D}). The parameters of the {\tt reflionx} model are the disc ionization parameter $\xi$, the iron abundance $A_{Fe}$, the temperature of the ionizing black body flux $kT_{refl}$ and a normalization $N_{refl}$. In order to account the relativistic effects, we used the convolution kernel {\tt relconv} \citep{2010MNRAS.409.1534D}. In this model, the emissivity of the disk is described as a broken powerlaw in radius (e.g., $\epsilon\propto r^{-q}$), giving three parameters: $q_{in}$, $q_{out}$ and $R_{break}$. Here we used an unbroken emissivity profile (fixed slope) by fixing $q_{out}=q_{in}$ (obviating the meaning of $R_{break}$) as the slope is not constrained by the data. The parameters of the {\tt relconv} model are the emissivity index $q$, the inner and outer disk radius $R_{in}$ and $R_{out}$, the disk inclination $i$ and the dimensionless spin parameter $a$. We set the emissivity to $q=3$, in agreement with a Newtonian geometry far from the NS \citep{2010ApJ...720..205C}. The spin frequency of the source Cyg~X--2 can be taken as $\sim 364$ Hz \citep{1998ApJ...493L..87W}. Following \citet{2000ApJ...531..447B}, the dimensionless spin parameter $a$ can be approximated as $a\simeq0.47/P_{ms}$ where $P_{ms}$ is the spin period in ms. Therefore, we fixed the spin parameter to $a=0.17$. We also fixed the outer radius $R_{out}=1000\;R_{G}$. \\ 

The addition of the relativistic reflection model improved the spectral fits significantly ($\chi^{2}/dof$=$1037/914$ and $\chi^{2}/dof$=$873/765$ for the non-dip and dip spectrum, respectively). The best-fit parameters are shown in Table~\ref{parameters}. The best-fit inner disk radius lies at $R_{in}=(2.5-6.0) R_{ISCO}$ for non-dip state and $R_{in}=(2.0-2.6) R_{ISCO}$ for the dip state. The inclination angle is found to be $i=23\pm2$ degree (consistent with \citealt{2010ApJ...720..205C}). The reflection component has an intermediate disc ionization of $\xi\simeq70-400$ erg s$^{-1}$ cm which is the typical range observed in both black hole and NS LMXBs ($\text{log}\xi\sim 2-3)$. The Fe abundance is less than that of solar composition ($\leq0.7$) for the non-dip spectrum and for the dip spectrum it was frozen to the value obtained from the non-dip spectrum as it was not well constrained. The fitted spectra (both non-dip and dip) along with the model components and the residuals are shown in Fig.~\ref{Fig7}.

\begin{table*}
   \centering
   \caption{ Best-fit non-dip and dip spectral parameters of the \nustar{} observation of the source Cyg~X--2 using model: {\tt TBabs$\times$(diskbb+compTT+relconv*reflionx)}. 1$\sigma$ error on all relevant parameters are quoted.} 
\begin{tabular}{|p{2.0cm}|p{4.5cm}|p{2.0cm}|p{2.0cm}|}
    
    \hline
    Component     & Parameter & \multicolumn{2}{c}{\nustar{} spectrum} \\
     \cline{3-4} 
                   &          & Non Dip & Dip \\ 
   \hline
    {\scshape tbabs}    & $N_{H}$($\times 10^{22} cm^{-2}$) &$0.22(f)$ & $0.22(f)$    \\
    {\scshape diskbb} & $kT_{in}$($\kev$) & $0.99_{-0.10}^{+0.18}$ & $1.66\pm0.05$  \\
    & $N_{diskbb}$[(km/10 kpc)$^{2}$cos$i$]   & $735_{-321}^{+223}$  &  $45_{-8}^{+10}$    \\

    {\scshape comptt} & $kT_{seed}(\kev)$ & $1.33\pm0.05$ & $0.76\pm0.01$    \\
    & $kT_{e}$ (keV) & $5.97_{-1.42}^{+7.25}$ &  $3.78_{-0.68}^{+1.61}$ \\
    & $\tau$ & $1.03_{-0.44}^{+0.40}$  &  $2.09_{-1.01}^{+0.93}$ \\
    
    & $n_{comptt}$~$^a$  & $0.16\pm0.14$ & $0.60_{-0.24}^{+0.19}$    \\ 
   {\scshape relconv} & $i$ (degrees) & $\leq 21 $ & $23\pm2$    \\
   & $R_{in}$($\times R_{ISCO}$) & $4.3\pm1.8$ & $2.0_{-0.26}^{+0.52}$ \\
   {\scshape reflionx} & $\xi$(erg cm s$^{-1}$) & $69_{-10}^{+30}$ & $390_{-130}^{+80}$ \\
   & $A_{Fe}$($\times$ Solar) &$0.5_{-0.4}^{+0.16}$  & $0.7(f)$ \\
   & $kT_{refl}$ (keV) & $2.45_{-0.07}^{+0.10} $ & $ 2.20\pm0.09  $ \\
   & $N_{refl}$~$^b$ & $11.6_{-4.2}^{+2.2}$ & $2.4_{-0.16}^{+0.30}$ \\
   & $F^{*}_{total}$ ($\times 10^{-8}$ ergs/s/cm$^2$) & $1.58 \pm 0.01$ & $0.95 \pm 0.01$ \\
   & $F_{diskbb}$ ($\times 10^{-8}$ ergs/s/cm$^2$) & $0.39 \pm 0.01$ & $0.35 \pm 0.01$ \\
   & $F_{comptt}$ ($\times 10^{-8}$ ergs/s/cm$^2$)& $1.07 \pm 0.01$ & $0.51 \pm 0.01$ \\
   & $F_{reflionx}$ ($\times 10^{-8}$ ergs/s/cm$^2$) & $0.12 \pm 0.01$ & $0.09 \pm 0.01$ \\
   & $L_{3-79 keV}$ ($\times 10^{38}$ ergs/s) & $1.21 \pm 0.01$ & $0.73 \pm 0.01$ \\	
  
\hline 
    & $\chi^{2}/dof$ &$1037/914$  & $873/765$   \\
    \hline
  \end{tabular}\label{parameters} \\
The outer radius of the {\tt relconv} spectral component was fixed to $1000\;R_{G}$ and the spin parameter was set to $a=0.17$ and $q=3$. ~$^{a,b}$ denotes the normalization component of the {\tt compTT} and {\tt relconv} model, respectively. Assumed a distance of 8 kpc and a mass of $1.5\;M_{\odot}$ for calculating the luminosity. $^{*}$All the fluxes are calculated in the energy band $3.0-79.0\kev$. 
\end{table*}

\begin{figure*}
  \psfrag{Non Dip}{\large Non Dip}
\psfrag{Dip}{\large Dip}
   \includegraphics[width=7.0cm, angle=0]{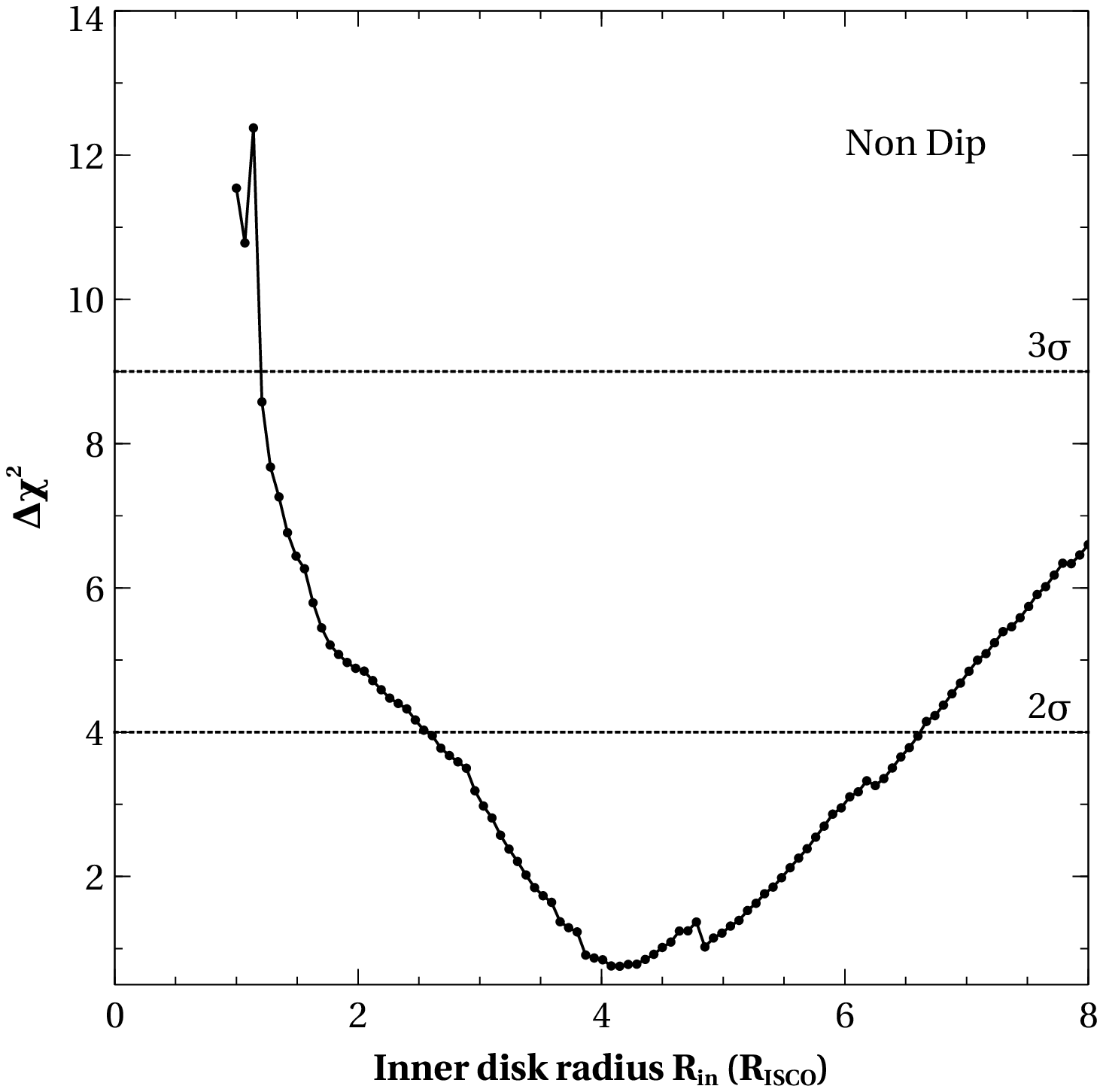}
   \includegraphics[width=7.0cm, angle=0]{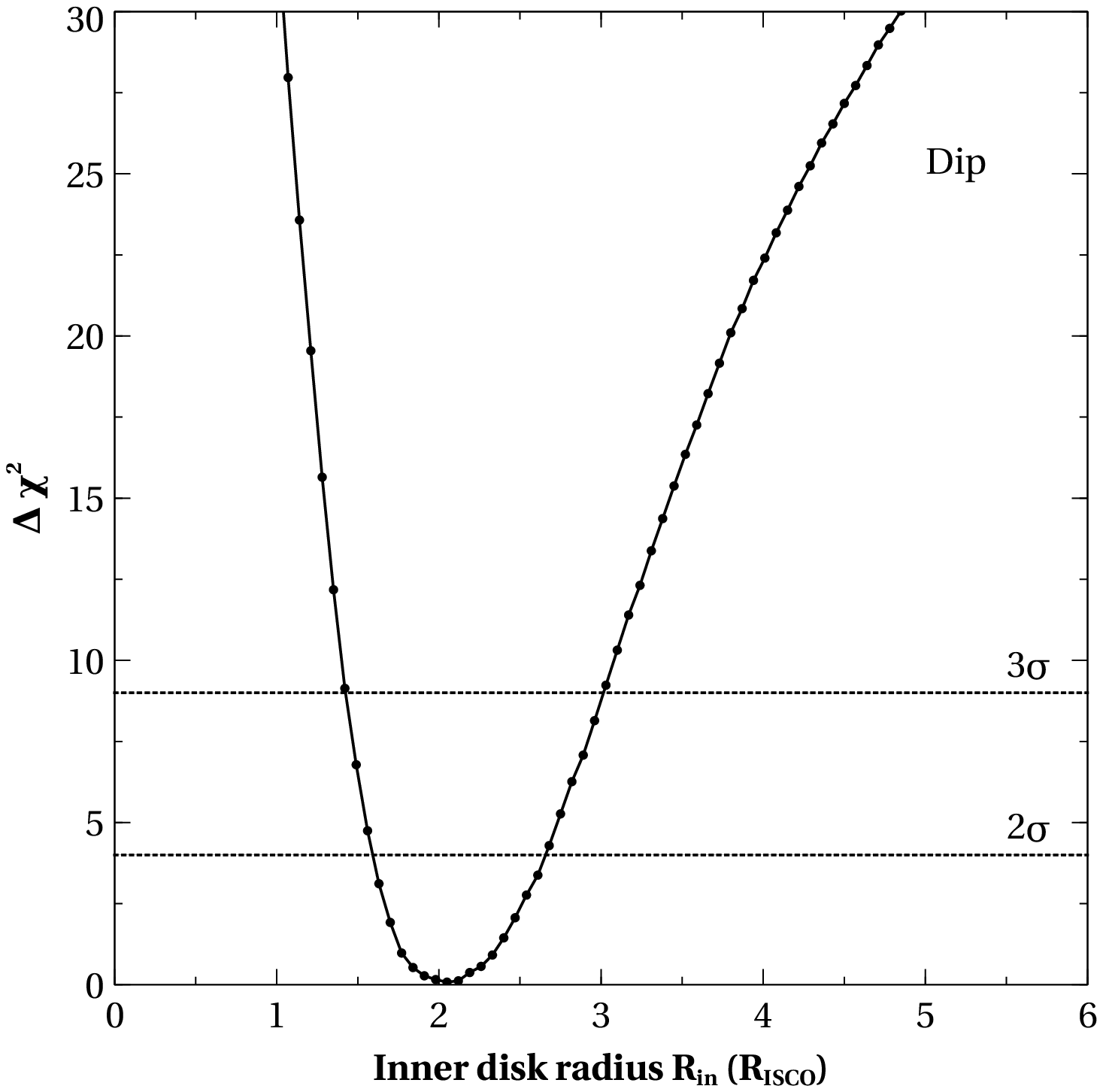}
   \caption{Both the panels shows the variation of $\Delta\chi^{2}(=\chi^{2}-\chi_{min}^{2})$ as a function of inner disc radius (in the unit of $R_{ISCO}$) obtained from the relativistic reflection model. We varied the inner disc radius as a free parameter upto $8\,R_{ISCO}$. Horizontal lines in both the panels indicate $2\sigma$ and $3\sigma$ significance level. } 
  \label{Fig8}
\end{figure*}
 
\begin{figure*}
   \centering
   \includegraphics[width=7.0cm, angle=0]{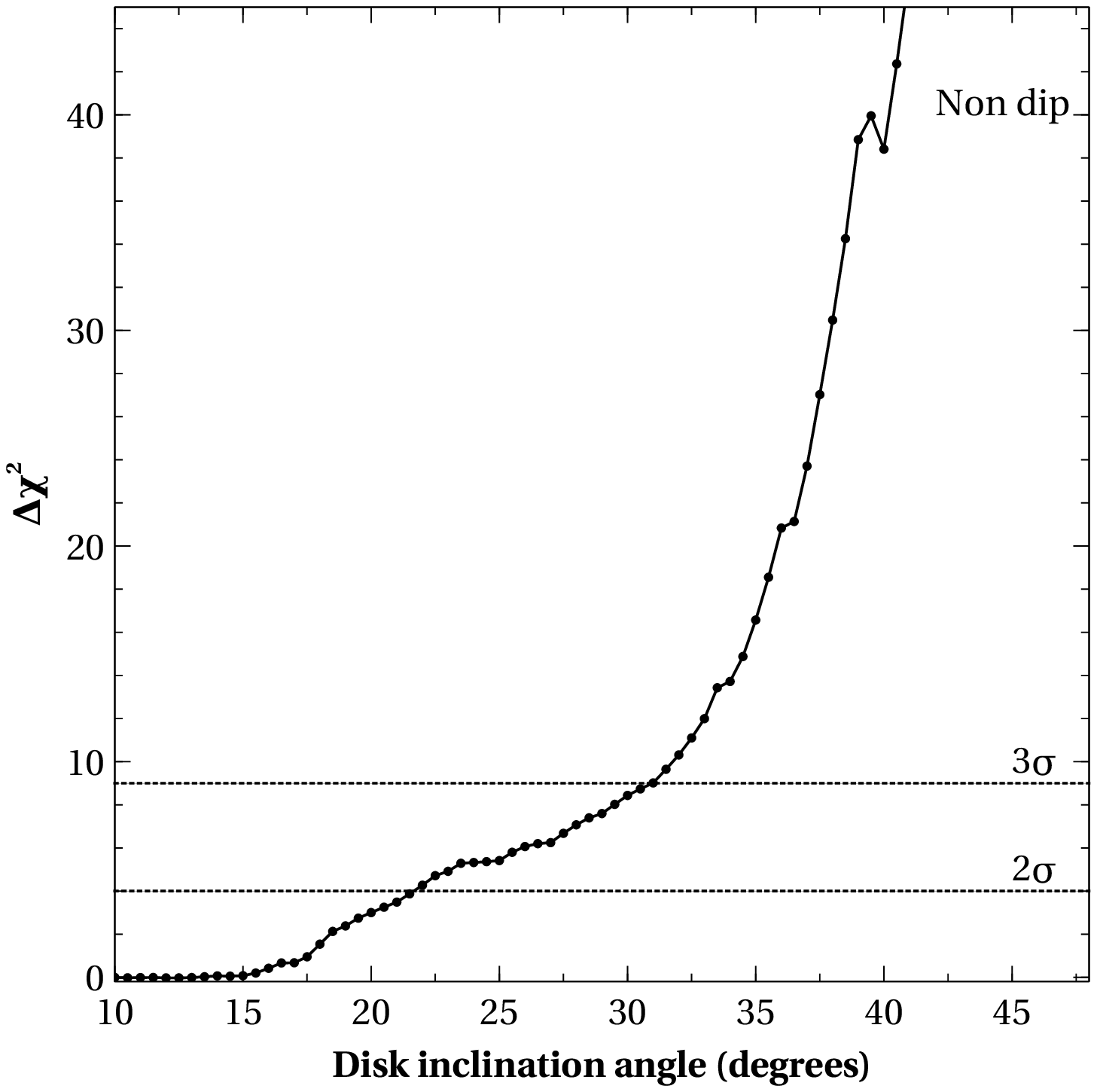}
   \includegraphics[width=7.0cm, angle=0]{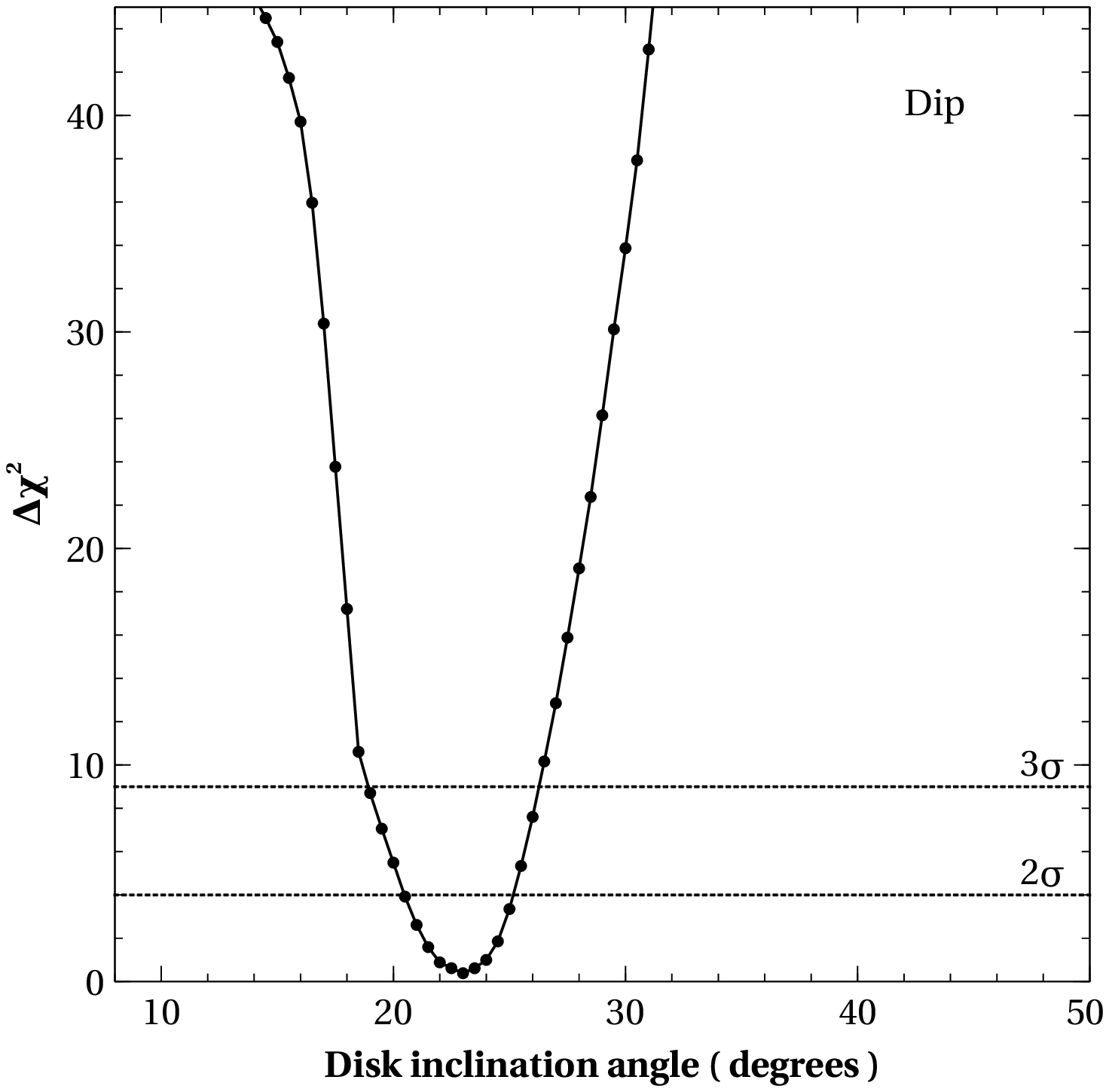}
   \caption{ Disk inclination angle delta chi-squared ($\Delta\chi^{2}$) distribution in the relativistic reflection model. We varied the disc inclination angle between 10 degree and 50 degree in both the states. Horizontal lines in both the panels indicate $2\sigma$ and $3\sigma$ significance level } 
   \label{Fig9}
   \end{figure*}

\section{Discussion}
We report on the \nustar{} observation of the bright Z-type low mass X-ray binary Cyg~X--2. From the hardness-intensity diagram (HID), it is confirmed that the source was in the so-called NB on the Z-track during this observation, although an extended FB is observed when the source move from non-dip to dip state. An increase in the fractional rms has been observed from the PDS as the source move from non-dip (NB) to dip (extended FB) state. Higher rms variability at long time scale is expected from the inner disc which moves closer to the NS surface. This is consistent with results from spectral analysis which show that as the source moves from non-dip to dip part, the inner disk radius decreases and the disk temperature increases. The broad-band $3-45\kev$ \nustar{} spectral data can be described by a continuum model consisting of a disk blackbody and thermal Comptonization. The source was in a soft spectral state with the $3-45\kev$ luminosity of $L\simeq(0.5-1.1)\times 10^{38}$ ergs s$^{-1}$, assuming a distance of 8 kpc. The spectral data required a significant reflection component, characterised by the broad Fe K$\alpha$ emission line. \\

From Fig.~\ref{Fig6}, it has been observed that there are differences between the Fe line profiles from non-dip and dip spectrum. During non-dip state, the Fe line profile is symmetric and there is a hump like structure at $15-30\kev$. While during dip state, Fe line profile is mostly asymmetric and $15-30\kev$ hump is significantly reduced. Symmetric line profile essentially indicate the existance of strong outflow/wind in the system which causes broadening of the blue wing of the line. However, such outflow is not present during dip state and we get an asymmetric line profile due to redshift only. Such scenario is further supported by the enhanced Comptonized flux during non-dip state ($\sim 20\%$ or so). Comptonization may occur also from the outflow and therefore it is increased during non-dip state. The enhanced Comptonization due to the presence of outflow/wind can provide stronger Compton back-scattering hump which is clearly visible in non-dip spectrum but faint in dip spectrum because of the absence of outflow/wind. The change in the shape of the HID is possibly due to the change in outflow properties (like wind on and off) which may affects the accretion geometry. This might be responsible for different seed photon temperature ($kT_{seed}$) in the non-dip and dip state for Comptinization. During this observation, a decrease in the hardness ratio (a factor of $\sim 1.6$ or so) has been observed when the source passes from the non-dip to the dip states. So, there is a softening of the spectrum during the dip state. This indicate that the transition from non-dip state to dip state is unlikely to be due to the absorption. From the final best-fit model, we found that the {\tt diskbb} component became more dominant from non-dip to dip state ($\sim 15\% $ increment in the {\tt diskbb} flux), while the comptonization component became weaker with decreasing cutoff energy, indicating a decrease in $kT_{e}$. The optical depth, $\tau$, is higher in the dip state compared to the non-dip state. \\

We fitted the reflection features with a relativistically blurred reflection model. We found that the line has an intermediate disc ionization ($\xi\sim70-400$ erg s$^{-1}$ cm) and a relatively low viewing angle ($i\sim23$ deg) but a large inner radius ($R_{in}=2.5-6.0\times R_{ISCO}$ from the non-dip spectrum and $R_{in}=2.0-2.6\times R_{ISCO}$ from the dip spectrum). Following \citet{2000arxt.confE...1V}, $R_{ISCO}$ can be approximated as $R_{ISCO}\simeq(6GM/c^{2})(1-0.54a)$. Thus for the spin parameter $a=0.17$ and $1.5\:M_{\odot}$ NS $R_{ISCO}=5.4\:GM/c^{2}$. So, the inner disk radius was located far from the NS surface at $R_{in}\simeq13.5-32.4\:GM/c^{2}$ ($\simeq30-73$ km), estimated from non-dip spectrum and at $R_{in}\simeq10.8-14\:GM/c^{2}$ ($\simeq24-32$ km) estimated from dip spectrum. Larger inner disk radius was also obtained by \citet{2002MNRAS.331..453D} after fitting \ginga{} data with the reflection model. \\

We compared the inner disk radius from the Fe line fitting with that implied from the {\tt diskbb} fits. The normalization component of the {\tt diskbb} is defined as $N_{diskbb}=(R_{in, diskbb}/D_{10})^{2}$cos$i$ (where $R_{in, diskbb}$ in km and $D_{10}$ is the distance in units of 10 kpc), which can be used as an important probe to constrain the inner radial extent of the accretion disk. Although the large uncertainties involving in factors like inner boundary assumptions \citep{1999MNRAS.309..496G}, spectral hardening \citep{2000MNRAS.313..193M}, the column density or inclination etc. hamper the measurement of the true inner-disk radii from {\tt diskbb} fits. From our best fit {\tt diskbb} normalization, we obtained an inner disk radius of $R_{in,diskbb}\simeq 17-26$ km  from the non-dip spectrum and $R_{in,diskbb}\simeq 7-9$ km from the dip spectrum for an inclination of $i=21-25$ deg and distance of 8 kpc. However, if we corrected it with the hardening factor $\simeq1.7$ (e.g. \citealt{2001ApJ...547L.119K, 2013ApJ...769...16R}), then it yielded $R_{in,diskbb}\simeq 49-74$ km from the non-dip spectrum and $R_{in,diskbb}\simeq 20-26$ km from the dip spectrum those are consistent with the location of the inner disk inferred from the reflection model in both spectra. In General, NS LMXBs exhibit a wide range of inferred inner disc radii, most sources show small inner disk radii of $6-15\:GM/c^{2}$ (see \citealt{2010ApJ...720..205C, 2015MNRAS.451L..85D, 2015MNRAS.449.2794D}), some sources show slightly larger inner radius of $15-30\:GM/c^{2}$ (see \citealt{2016A&A...596A..21I, 2011ApJ...731L...7M, 2013MNRAS.429.3411P, 2016ApJ...819L..29K}). At the same time, a significantly larger truncation radii of $\geq 80GM/c^{2}$ has been observed for two NS LMXBs (\citealt{2016MNRAS.456.4256D, 2014ApJ...796L...9D}). In this observation of the source Cyg~X--2, we have observed a moderate disc truncation of inner disc radii $\simeq11-32\:GM/c^{2}$. If it is assumed that the truncation of the disc is caused by the magnetosphere, it would imply a large NS magnetic field ($\geq 4\times 10^{8}$ G).  \\

We can assume that the inner accretion disk is truncated at the magnetospheric radius rather than its surface. At the magnetospheric radius, magnetic pressure balances the ram pressure from the infalling material. From the source luminosity and reasonable assumption about mass and radius, we can estimate the magnetic field strength. We used the following equation of \citet{2009ApJ...694L..21C} which was a modified version of the formulation of \citet{2009MNRAS.400..492I} to calculate the magnetic dipole moment. \\
\begin{equation}
\begin{split}
\mu=&3.5\times 10^{23}k_{A}^{-7/4} x^{7/4} \left(\frac{M}{1.4 M_{\odot}}\right)^{2}\\
 &\times\left(\frac{f_{ang}}{\eta}\frac{F_{bol}}{10^{-9} \text{erg}\: \text{cm}^{-2}\: \text{s}^{-1}}\right)^{1/2}
 \frac{D}{3.5\: \text{kpc}} \text{G}\; \text{cm}^{3}
\end{split} 
\end{equation}
where $\eta$ is the accretion efficiency in the Schwarzschild metric, $f_{ang}$ is the anisotropy correction factor. The coefficient $k_{A}$ depends on the conversion from spherical to disk accretion (numerical simulation suggest $k_{A}=0.5$ whereas theoretical model predict $k_{A}<1.1$). \citet{2009ApJ...694L..21C} modified $R_{in}$ as $R_{in}=x\:GM/c^{2}$. We estimated flux in the $0.01-100\kev$ range is of $F_{bol}=2.2\times10^{-8}$ erg cm$^{-2}$ s$^{-1}$. We assumed $D=8$ kpc, $M=1.5M_{\odot}$ and $R=10$ km. Using $R_{in}\leq32 R_{G}$ from the \nustar{} Fe line fit, along with assuming  $k_{A}=1$, $f_{ang}=1$ and $\eta=0.1$, leads to magnetic field strength of $B\leq7.6\times10^{9}$ G at the magnetic poles. This value is consistent with \citealt{1996ApJ...473L..45W} ($\sim8.5\times10^{9}$ G) when the source enters in the normal branch of the Z track. However, if we assume $k_{A}=0.5$, the magnetic field strength is increased nearly by a factor of 3.\\

We calculated the Keplerian frequency associated with the truncation radius using the following relation by \citet{2000arxt.confE...1V}
\begin{equation}
\nu_{orb}\approx1200(r_{orb}/15 \text{km})^{-3/2} m_{1.4}^{1/2}\; \text{Hz}
\end{equation}
which was found to be in the range $155-611$ Hz. This frequency range is consistent with the spin frequency of Cyg~X--2 if the difference between kHz QPO of $\sim364$ Hz is taken as a spin estimate \citep{1998ApJ...493L..87W}. \\

\section{summary}
In this paper, we have analyzed a \nustar{} observation of the Z-type NS LMXB Cyg~X--2. Our findings obtained from the timing and the spectral studies may be summarized as follows -- 
\begin{itemize}
\item
The source exhibited a sudden decrease in the count rate (dips) and stronger variability in $3-79\kev$ lightcurve.
\item
It is confirmed from the HID that the source was in the so-called NB on the Z-track during this observation, although an extended FB is observed when the source moves from non-dip to dip state.
\item
An increase in the fractional rms has been observed from the PDS as the source move from the non-dip (NB) to the dip (extended FB) state.
\item
The broad band $3-45\kev$ \nustar{} spectral data can be described by a continuum model consisting of a disk blackbody and thermal Comptonization.
\item
The source was found in a soft spectral state with the $3-45\kev$ luminosity of $L\simeq(0.5-1.1)\times 10^{38}$ ergs s$^{-1}$, assuming a distance of 8 Kpc.
\item
We found broad, relativistic Fe K$\alpha$ emission line and a hump like structure at $15-30\kev$. From the relativistic reflection model fitting, we were able to measure the inner radius of the accretion disc.
\item
The Keplerian frequency associated with the truncation radius was found to be in the range $155-611$ Hz.
\item
We estimated an upper limit of the magnetic field strength of $\leq7.6\times10^{9}$ G at the magnetic poles with an assumption that the disc is truncated at the magnetospheric radius.
\end{itemize}

\section{Acknowledgements}
This research has made use of data and/or software provided by the High Energy Astrophysics Science Archive Research Centre (HEASARC). Aditya S. Mondal would like to thank Inter-University Centre for Astronomy and Astrophysics (IUCAA), Pune, India for hosting him during his visits. BR likes to thank IUCAA for hospitality and other facilities extented to him during his visits under the Visiting Associateship programme.

\def\apj{ApJ}
\def\apjl{ApJl}
\def\pasp{PASP} \def\mnras{MNRAS} \def\aap{A\&A} \def\physerp{PhR} \def\apjs{ApJS} \def\pasa{PASA}
\def\pasj{PASJ} \def\nat{Nature} \def\memsai{MmSAI} \def\araa{ARAA} \def\iaucirc{IAUC} \def\aj{AJ} \def\aaps{A\&AS} \def\ssr{SSRv}
\bibliographystyle{mn2e}
\bibliography{cyg_x2}

\end{document}